\documentclass[a4paper]{jpconf}
\newcommand{\comm}[1]{\ensuremath{[#1]}}

\usepackage[latin1]{inputenc}  

\usepackage{poormath}
\usepackage{amssymb}

\usepackage{defs}
\newcommand{\nicefrac}[2]{#1/#2}
\newcommand{\form}[1]{\ensuremath{_{[#1]}}}

\usepackage{braket}
\begin{document}

\begin{flushright}
 CPHT-C050.0606\\
hep-th/0607027
 
\end{flushright}

\title{Corfu 05 lectures - part I:\\ Strings on curved backgrounds}

\author{Domenico Orlando$^{\spadesuit \heartsuit }$, P.Marios Petropoulos$^\spadesuit $}

\address{$^\spadesuit $ Centre de Physique Théorique, École
polytechnique\footnote{Unité
      mixte du CNRS et de l'École polytechnique, UMR 7644.} \\
    91128 Palaiseau, France } 
\address{$^\heartsuit $ Theoretische Natuurkunde, Vrije Universiteit Brussel and\\
    The
    International Solvay Institutes\\
    Pleinlaan 2, B-1050 Brussels, Belgium}
  
\ead{domenico@cpht.polytechnique.fr, marios@cpht.polytechnique.fr}
\begin{abstract}

In these introductory lectures we summarize some basic facts and
techniques about perturbative string theory (sections 1 to 6).
These are further developed (sections 7 and 8) for describing
string propagation in the presence of gravitational or gauge
fields. We also remind some solutions of the string equations of
motion, which correspond to remarkable (NS or D) brane
configurations.

A part II by Emilian Dudas will be devoted to orientifold
constructions and applications to string model building.

\end{abstract}

\section{Fields versus strings}
\label{chap:intro}

Field theories contain in general an
arbitrary number of particles (fields) and their mutual interactions
are essentially constrained only by the requirements of
renormalizability and unitarity. This results in a large freedom in
choosing these interactions (Yang--Mills, Yukawa, $\phi^4$ just to name
some of the best studied) that is in practice only limited by
phenomenological constraints. Enough freedom remains to incorporate
some luxury items such as grand-unified groups, supersymmetry or
Kaluza--Klein spectra at least if they do not contradict the available
experimental data. This is nevertheless not enough to introduce
gravity in the picture, at least unless we choose to abandon some
commonly accepted rules.

A possible
way out (and possibly the only really promising one at this time)
is provided by string theory. In this case an infinite spectrum of
particles arises but this time naturally arranged in
representations of a superalgebra and defined in dimension $D>4$.
This implies the presence of a Kaluza--Klein spectrum. Among the
highlights there is the fact that -- and this is different from
the field case -- gauge interactions are not added \emph{ad hoc}
but do appear naturally, and together with them gravitational
interactions, in the form of supergravity. Moreover the presence
of large gauge groups is in some way natural.

Of course many open questions still remain. Among them
the fact that the little freedom allowed by the consistency
constraints in ten dimensions becomes pretty large in four.  Other
still open problems concern the breaking of supersymmetry or the
role that string theory can have in explaining the cosmological
evolution. An important feature (or, a weakness, depending on the
point of view) is the fact that string theory incorporates more
exotic objects. In particular the theory is only
consistent if we add extended objects, like D$p$~branes or
NS5~branes and higher-order forms. The latter are generalization
of the field-theory gauge fields and appear in the Ramond--Ramond
sector of the type IIA and IIB theories where
they are coupled to D$p$~branes\footnote{In the same way as the
  electromagnetic field, which is a two-form, is electrically coupled
  to a point particle, a $(p+2)$-form is electrically coupled to a
  D$p$~brane} and in the universal sector in terms of a three-form
$H_{[3]}$ that is electrically coupled to the string and
magnetically to the NS5~brane. Although originally designed around
flat space, these branes can propagate in curved (gravitational)
backgrounds with various gauge fields switched on (fluxes through
the compact sector of the geometry) and have become an almost
indispensable tool for probing aspects such as string gravity,
black hole (thermo)dynamics, supersymmetry breaking, moduli
stabilization, in analyzing holography and decompactification
regimes and -- last but not least -- in the quest for
time-dependent solutions with a cosmological interest.

A fact -- at the root of many of the present complications -- is
that strings in general prefer spheres $S^n$ and anti-de Sitter
spaces $\mathrm{AdS}_n$ (it suffices to recall the partial
breaking of supersymmetry or holography, that naturally bridges
strings and super-Yang--Mills theories) while do not seem to like
hyperbolic planes $H_n$ or de Sitter spaces $\mathrm{dS}_n$ --
which is a sort of archetype of the difficulties that one
encounters while trying to make contact with cosmology.

The purpose of the present lecture notes is to summarize some of the
basic tools, necessary for addressing the latter issues.
Sections~\ref{cha:string-motion-d} and~\ref{cha:advanc-string-moti-1}
deal with the basics on perturbative closed strings which are extended
to open strings in Sec.~\ref{cha:open-strings-1}. Once the massless
string spectra are established, coherent states of massless
excitations can be generated, which act as classical backgrounds for
the corresponding fields (section~\ref{cha:strings-backgr-field}).
String propagation in such environments is further analyzed in
section~\ref{cha:remarkable-solutions}.

These notes are elementary and many important issues are missing.
For this reason we avoid referring to the original papers in the
course of the main text. A list of references for further reading
is made available at the end\footnote{Further reviews on string
theory
  can be found at the following URL
  \texttt{http://www.slac.stanford.edu/spires/find/hep/www?hn=pre+reviewIf}}.


\section{String motion in $D$-dimensional flat spacetime}
\label{cha:string-motion-d}

\subsection{Free-falling relativistic particle}

Let us start from the most simple system: the motion of a free-falling
relativistic point particle with mass $m \neq 0$. The manifest
relativistic action is just given by the length of the trajectory between the
two extremal fixed points in spacetime:
\begin{equation}
S = - m \int_\mathrm{i}^\mathrm{f} \di \tau \: \sqrt{- \eta_{\mu
\nu } \dot x^\mu \dot x^\nu} = - m \int_\mathrm{i}^\mathrm{f} \di
s .
\end{equation}
Since the Compton wavelength is defined as $\lambda = \nicefrac{\hbar}{mv}$ we
expect a classical behaviour for large $m$ and quantum behaviour for
small $m$, where ``large'' and ``small'' are defined with respect to some
energy scale of the system.

As it stands this system is overdetermined. In fact, in
a Hamiltonian formalism (not manifestly relativistic) the motion
$\mathbf{x} = \mathbf{x} (\tau)$ can be described in a $2 \left( D
- 1 \right) $-dimensional phase space defined by:
\begin{equation}
  \begin{cases}
    \tau = x^0 \\
    p^\mu p_\mu = - m^2.
  \end{cases}
\end{equation}
This is the reason why, apart from being Poincaré invariant, the
action is also reparameterization-invariant. This local symmetry can
be used to remove one degree of freedom by imposing $\eta_{\mu \nu } \dot x^\mu
\dot x^\nu = -1 $ (that is equivalent to $p^\mu p_\mu = - m^2$) and still we
will be left with the residual time translation invariance $\tau \to \tau + a$
which allows to remove the initial value for the time coordinate.

The presence of redundant (spurious) degrees of freedom is a
common trait of manifest relativistic invariant theories. The
standard example is given by the electromagnetic potential
four-form $A_\mu$ which contains two unphysical polarizations as a
consequence of the gauge invariance $A_\mu \to A_\mu +
\partial_\mu \Lambda$. They can be removed, \emph{e.g.} by going
to the light-cone gauge, but the price to pay is the loss of
manifest Lorentz invariance.

\subsection{Strings}

Let us now move to strings. The natural extension of the previous
description is provided by the Nambu--Goto action measuring the
area of the world-sheet swept by the string in its motion from an
initial configuration~$i$ to a final configuration~$f$:
\begin{equation}
  \label{eq:Nambu-Gotto}
  S = -T \int_\mathrm{i}^\mathrm{f} \di^2 \zeta \: \sqrt{ - \det ( \hat g )} =
  - T \int_\mathrm{i}^\mathrm{f} \di A,
\end{equation}
where $\hat g$ is the pull-back of the spacetime metric on the
world-sheet (in components $\hat g_{\alpha \beta} =
\partial_\alpha X^\mu \partial_\beta X^\nu \eta_{\mu \nu }$),
parameterized by the variables $\zeta^\alpha, \alpha=1,2$.
The parameter $T$ is the
string tension, whose length is not fixed but depends on the
configuration. In average this is given by $\ell \sim
\nicefrac{1}{\sqrt{T}} = \sqrt{2 \pi \alpha^\prime}$.  This new
parameter $\alpha^\prime$ fixes the scale and hence the behaviour
of the string: classical when $\alpha^\prime $ is small and
quantum when $\alpha^\prime $ is large with respect to the typical
energies of the system.  Just to have an order of magnitude,
$\alpha^\prime $ is of the order of
$\nicefrac{1}{\left(10^{19}\mathrm{GeV}\right)^2}$ (the inverse
square of the Planck mass). For this reason it can be used as the
``small parameter'' in a perturbative expansion. We will
distinguish between \emph{exact solutions}, known to all orders in
$\alpha^\prime$, including non-perturbative effects and
\emph{approximate
  solutions} for which, in general, only the first terms of the
development are known. The other possible expansion parameter
$g_{\text{st}}$, controlling the string topological expansion, will be
introduced in the next chapter.

The action
(\ref{eq:Nambu-Gotto}) is Poincaré invariant and it is not
difficult to show its reparameterization invariance symmetry
$\zeta^\alpha \to \tilde \zeta^\alpha ( \zeta^\beta )$. The latter
plays a fundamental role in string theory at many levels,
including some aspects that we do not yet fully understand
(\emph{e.g.} the full spacetime gauge invariance which is a guide
for understanding the issues of second quantization). Just as
before this local symmetry (or, equivalently, the system being
overconstrained) introduces a redundancy in the degrees of freedom
where two coordinates can be used as world-sheet parameters. Of
course a light-cone-like gauge can be used in which the physical
degrees of freedom are the only ones appearing again at the
expense of the manifest Lorentz invariance.

The
classical action is just the first step: although canonical
quantization of the string can be performed successfully in this
formalism, the square root poses some problems in the path-integral
quantization and in the generalization to superspace. This is why
Brink, Di Vecchia, Howe, Deser, Zumino introduced the Polyakov action
(named after the latter for having analyzed its path-integral
quantization). The approach is reminiscent of Lagrange multipliers,
\emph{i.e.} consists in introducing a new auxiliary field and a new
local symmetry in order to have a classically equivalent dynamics.
This action reads:
\begin{equation}
\label{eq:polyakov-flat}
  S = - \frac{1}{4 \pi \alpha^\prime } \int \di^2 \zeta \: \sqrt{-h} h^{\alpha
    \beta } \partial_\alpha X^\mu \partial_\beta X^\nu \eta_{\mu \nu } -
  \frac{\lambda }{4 \pi } \int \di^2 \zeta \: \sqrt{-h} R
\end{equation}
where $h$ is the extrinsic metric of the world-sheet and $R$ the
scalar curvature.  The reparameterization invariance is supplemented with
the Weyl invariance $h_{\alpha \beta } \to \mathrm{e}^{2 \rho}
h_{\alpha \beta}$. The Einstein--Hilbert action in two dimensions
is purely topological: this is why $h$ is non-dynamical and can
hence be locally set to $h_{\alpha \beta } = \delta_{\alpha \beta}$.\\
In terms of the equations of motion, the Weyl invariance is
equivalent to the constraint
\begin{equation}
  \partial_\alpha X^\mu \partial_\beta X^\nu \eta_{\mu \nu} - \frac{1}{2}
  \eta_{\alpha \beta } \left( \partial_\gamma X^\mu \partial^\gamma X^\nu \right)
  \eta_{\mu \nu} = T_{\alpha \beta } = 0
\end{equation}
that must be satisfied by a physically sensible solution to the
equations of motion. In the usual notation one introduces the
variables $\left( \sigma, \tau \right) = \zeta^\alpha$ and the
left- and right-moving light-cone parameters $\zeta^± = \tau ±
\sigma $. In these terms the equations of motion read $\partial_+
\partial_- X^\mu = 0$ and the general solution can be easily cast
in a Fourier-expansion form as the sum of the center of mass
motion and a series of harmonics:
\begin{equation}
\label{eq:Fourier-expansion}
  X^\mu (\sigma, \tau ) = x_0^\mu + 2 \alpha^\prime p^\mu \tau +
  \begin{cases}
    \displaystyle{\sum_{n \neq 0}} \imath \sqrt{\nicefrac{\alpha^\prime}{2}} \frac{1}{n}
    \alpha_n^\mu \mathrm{e}^{-2 i n \left( \tau - \sigma \right) }  \text{(right
      sector)} \\ \displaystyle{\sum_{n \neq 0}} \imath \sqrt{\nicefrac{\alpha^\prime}{2}}
    \frac{1}{n} \tilde \alpha_n^\mu
    \mathrm{e}^{-2 i n \left( \tau + \sigma \right) }  \text{(left sector)} \\
  \end{cases}
\end{equation}
with the conditions $\alpha^\mu_{-n} = \left( \alpha^\mu_n \right)^\ast $ and $\tilde
\alpha^\mu_{-n} = \left( \tilde \alpha^\mu_n \right)^\ast $.  The constraints $T_{\alpha \beta}$
translate into constraints over $p^\mu, \alpha^\mu_n, \tilde \alpha^\mu_n$ that are
therefore not independent. One possible choice consists in eliminating
two light-cone directions $X^± $ which are then expressed as functions
of the others. As a result $x_{0}^±,x_{0}^i, p^±,p^i$ and
$\alpha^i_{n},\tilde{\alpha}\i_{n}$ remain independent modulo the two
(mass-shell and level-matching) constraints:
\begin{equation}
  M^2_{\text{string}} \equiv - p_\mu p^\mu = \frac{4}{\alpha^\prime}
  \sum_{n=1}^\infty \alpha^i_{-n} \alpha^j_n \delta_{ij} = \frac{4}{\alpha^\prime}
  \sum_{n=1}^\infty \tilde \alpha^i_{-n} \tilde \alpha^j_n
  \delta_{ij}.
\end{equation}
The latter expression defines the mass squared of the string in a
given configuration. The classical mass spectrum of the bosonic
string is a continuum above a zero-mass ground state.


\section{Basics on string quantization in $D$-dimensional flat
  spacetime}
\label{sec:basics}

Different options exist for quantizing constrained systems with
gauge symmetries. These include:
\begin{itemize}
\item light-cone quantization, which consists in considering only the
  physical degrees of freedom. This automatically guarantees
  unitarity but Lorentz invariance is not
  manifest;
  \item covariant quantization in which all the degrees of freedom are taken
  into account. Consistency conditions must be imposed, and
  hopefully remove negative-norm states from the spectrum.
\end{itemize}
Other possible choices include path integral quantization, \textsc{brst}
quantization, etc. In this chapter we will concentrate on the
light-cone canonical and the path-integral quantizations.

\subsection{Canonical quantization}

Let us consider the light-cone (canonical) quantization starting
from the Fourier expansion in Eq.~\eqref{eq:Fourier-expansion}.
The zero-modes satisfy the usual commutation relations:
\begin{equation}
  \comm{ x^\mu, p^\nu } = \imath \eta^{\mu \nu} ,
\end{equation}
and $\alpha^{\dagger \mu}_{n>0}, \tilde \alpha^{\dagger
\mu}_{n>0}$ have the roles of creation operators. A general state
is written as:
\begin{equation}
  \ket{\mathbf{p}, i_r, m_r, \ldots, j_s, n_s, \ldots } = \left( \alpha^{i_r}_{m_r}
    \ldots \right)  \left( \tilde \alpha^{j_s}_{n_s} \ldots\right) \ket{ \mathbf{p} } ,
\end{equation}
where $i_r, j_s = 1, 2, \ldots, D-1$. The left and right levels
are
\begin{align}
  N = \sum_r m_r , && \bar N = \sum_s n_s ,
\end{align}
so that the level-matching and mass-shell conditions read:
\begin{gather}
  N = \bar N , \\
  -p^2 = M^2 = \frac{4}{\alpha^\prime} \left( N - \frac{D-2}{24}
  \right) ,
\end{gather}
where the $(D-2)/24$ is the quantum two-dimensional vacuum energy.
It is useful to point out that each state is a one-particle state
in some representation of the Poincar\'e group. Hence, the Hilbert
space spans an infinity of different Poincar\'e representations,
massless and massive with a mass of the order of the Planck mass
($10^{19}\:\mathrm{GeV}$). First-quantized strings are described
in terms of a second-quantized field theory. This is at the moment
the state of the art and no satisfactory extension is yet known.
In some sense this is the same situation in which the interaction
between protons and electrons was before the introduction of
\textsc{qed}: propagators, vertices and a collection of ad-hoc
perturbation rules. Actually, string theory is in this respect
more satisfactory since it is very constrained (practically all
aspects are frozen, even the number of dimensions), including the
way in which string states (the equivalent of particles) interact.

Let us now give a closer look at the spectrum:
\begin{itemize}
\item the ground level
  $\ket{\mathbf{p}}$ has mass:
  \begin{equation}
    - p^2 = M^2 = -\frac{D-2}{24} ,
  \end{equation}
  which is tachyonic (and hence unstable\footnote{The presence of
    tachyons usually means that the theory is studied around a false vacuum.
    Theories with tachyons have usually been discarded
    because of the lack of a non-perturbative description, but the
    situation is lately changing.}) unless $D=2$;
\item the first level is:
  \begin{equation}
    \ket{\mathbf{p}, 1,i, 1,j} = \alpha^i_{-1} \tilde \alpha^j_{-1}
    \ket{\mathbf{p}}
  \end{equation}
  and has mass
  \begin{equation}
  -p^2 = M^2 = 1 - \frac{D-2}{24}.
  \end{equation}
  This is a symmetric tensor with $\left( D - 2 \right) \times \left( D - 2
  \right)$ degrees of freedom which we can decompose in its traceless
  symmetric part ($D \left( D -3 \right)/2$ \emph{dof}), antisymmetric
  ($\left(D -2 \right)\left( D - 3 \right) /2$) and trace ($1$
  \emph{dof})\footnote{These three components will be later identified
    with the graviton, the Kalb--Ramond field and the dilaton. Such an
    identification requires the treatment of interactions.}. Since only
  the transverse degrees of freedom appear here, we obtain a
  representation of the Poincar\'e group if the particle is massless,
  that is if $D = 26 $.  This tensor is part of the universal sector
  of string theory, \emph{i.e.} it appears in the massless sector of
  every model;
\item higher levels are massive, bosonic representations.
\end{itemize}

Let us pause for a moment and
discuss this latter result. Bosonic string theory naturally
contains a critical dimension $D = 26 $. This can be interpreted
in various ways. In the light-cone quantization on can show that
the conserved charges associated to the Lorentz currents do not
close if $D \neq D_{\text{cr}}$, \emph{i.e.} there is an anomaly:
the Lorentz algebra is not only non-manifest but not present
altogether. In a covariant quantization scheme the critical
dimension is required for unitarity. The path integral
quantization allows for a different point of view on the critical
dimension and brings extra information about the perturbative
expansion and the dynamics.

\subsection[Path integral quantization]{Path integral quantization: the
  string perturbative expansion}

In the path integral quantization, the partition function
-- or any correlator -- is written as an integral over the
embedding coordinates $X^\mu$ and the non-dynamical
two-dimensional metric $h_{\alpha \beta}$:
\begin{equation}
  \label{eq:Z-Polyakov}
  Z = \int \frac{\mathcal{D} X^\mu \mathcal{D} h_{\alpha \beta}
    }{\text{[Volume of diffeomorphism and Weyl groups]}} \e^{-S[X^\mu,
    h_{\alpha \beta}]} .
\end{equation}
The minus sign in the exponential is consequence of the Polyakov
prescription for a Wick rotation on the world-sheet. Out of this we can derive
important results about the string loop expansion and the critical dimension.

\subsection{String loop expansion}
\label{sec:string-loop}

The integral over $h_{\alpha \beta}$ can be decomposed as a sum
over the topologies of the two-dimensional world-sheet and an
integral over the metrics with fixed topology:
\begin{equation}
  \int \mathcal{D} h_{\alpha \beta} = \sum_{\text{topologies}} \e^{-\lambda \chi}
  \int_{\text{fixed topology}} \mathcal{D} h_{\alpha \beta} ,
\end{equation}
where $\chi$ is the Euler number
\begin{equation}
  \chi = \frac{1}{4\pi} \int \di^2 \zeta \: \sqrt{-h} R = 2 - 2 \gamma - M ,
\end{equation}
$\gamma $ being the genus and $M$ the number of boundaries. An important
consequence of this analysis is the appearance of a \emph{unique}
string vertex with coupling constant
\begin{equation}
  \e^\lambda = g_{\text{closed string}}.
\end{equation}

The various features that have emerged sofar exhibit major
differences with ordinary field theory:
\begin{itemize}
  \item the spectrum of particle, \emph{i.e.} the string states belonging to a
  given Poincar\'e representation, is given once and forever;
\item the interactions are fixed by the Polyakov path integral
  quantization;
\item the string coupling is (at least in principle) dynamically
  fixed: $\lambda$ turns out to be the vacuum expectation value of the dilaton;
\item each string state corresponds to a \emph{vertex operator},
  \emph{i.e.} an operator of the two-dimensional theory that we insert
  to compute the amplitudes;
\item even if a non-perturbative formulation is not available, we
  still have a handle on non-perturbative effects with respect to
  $g_{\text{st}}$. With this we mean that in some situation we can
  use dualities to map a problem in the $g_{\text{st}} \gg  1 $ regime
  to another problem with coupling $1/g_{\text{st}} \ll 1$ which can
  then be tackled with perturbative techniques.
\end{itemize}

A last remark is in order: the very concept of world-sheet is by
definition semiclassical and breaks down in some regimes,
controlled by $g_{\text{st}}$. At this moment we still miss an
appropriate  non-perturbative description from which the Polyakov
prescription would follow.

\subsection{Critical dimension: critical versus non-critical strings}
\label{sec:critical-dimension}

In the
approach of canonical light-cone quantization the critical dimension
$D = 26$ appears to guarantee that transverse-two-tensor modes are
massless. A different interpretation for this phenomenon, is available
from the path-integral viewpoint. The measure appearing in the
integral of Eq.~(\ref{eq:Z-Polyakov}) is not Weyl invariant: there is
a quantum anomaly which is an obstruction to the decoupling of the
two-dimensional metric, and is reflected in the central charge of the
two-dimensional \textsc{cft}. Two different consistent string regimes
finally emerge:
\begin{itemize}
\item for $D = 26 $ (critical strings), $h_{\alpha \beta }$ is not dynamical;
\item for $D < 26 $, the scale factor of $h_{\alpha \beta}$ contributes to the
  spectrum as the Liouville mode.
\end{itemize}
Liouville theory was studied in the early eighties as a
two-dimensional field theory. It reappeared in the nineties in the
developments of \emph{matrix models} and non-perturbative
two-dimensional quantum gravity and has again attracted some
attention recently in the framework of holography.


\section{Advanced string motion: extra degrees of freedom and open~strings}
\label{cha:advanc-string-moti-1}

\subsection{Fermionic degrees of freedom}

The states we built in Sec.~\ref{cha:string-motion-d} are
by construction only integer-spin representations of the Poincar\'e
algebra, which implies that in order to incorporate fermions we need
to add extra degrees of freedom. This can be done following two types
of formalisms:
\begin{itemize}
\item the \emph{Green--Schwarz} where spacetime spinors are introduced
  with manifest spacetime supersymmetry. This formalism is heavy; it
  turns out to be convenient only in some specific situations.
\item the \emph{Neveu--Schwarz--Ramond} which can be used more
  generally, has no explicit spacetime supersymmetry and does not allow
  a general treatment of the Ramond--Ramond fields in terms of
  sigma-model.
\end{itemize}

Let us introduce the \emph{fermionic coordinates} $\psi^\mu$ and a
sort of superspace described by couples $\left( x^\mu, \psi^\mu
\right)$. The string motion is captured by a set of functions:
\begin{equation}
  \begin{cases}
    x^\mu = X^\mu ( \zeta ), \\
    \psi^\mu = \Psi^\mu ( \zeta ) .
  \end{cases}
\end{equation}
The two-dimensional Majorana spinors appear as world-sheet fields
and we need to specify their dynamics\footnote{It is worth to
point out that introducing
  world-sheet fermions does not automatically guarantee the existence
  of spacetime spinors.}. The
natural choice consists in adding to the action
(Eq.~\eqref{eq:polyakov-flat}) an ordinary Dirac massless
 term:
\begin{equation}
  S = - \frac{1}{4 \pi \alpha^\prime} \int \di^2 \zeta \: \sqrt{-h} \left(
  h^{\alpha \beta} \partial_\alpha X^\mu \partial_\beta X^\nu - \imath \bar
  \psi^\mu \rho^a \e^{\alpha}_{\phantom{\alpha}a} D_\alpha \psi^\nu \right)
  \eta_{\mu\nu} ,
\end{equation}
where $h$ is the two-dimensional metric, $\rho^a$ the
two-dimensional Dirac $\gamma$ matrices and
$\e^{\alpha}_{\phantom{\alpha}a}$ the \emph{zwei-bein}. This action
has the following symmetries: local Weyl-rescaling of the
two-dimensional metric, two-dimensional diffeomorphism invariance,
global two-dimensional $N = ( 1, 1 )$ supersymmetry and global
spacetime Poincar\'e invariance.

In spite of its high symmetry, the action at hand is incomplete.
This caveat emerges during  the quantization. In fact, using the
canonical mode expansion for $\psi^\mu$ and imposing the
anticommutation relations one witnesses the appearance of
negative-norm states which would decouple if we could go to a
light-cone gauge and eliminate say \emph{e.g.} $\Psi^0$ and
$\Psi^{D - 1}$ just as we did for $X^0$ and $X^{D-1}$. This is
possible in the bosonic sector thanks to the Weyl and \emph{diff}
local invariance which for the action above do not have any
counterpart in the fermionic sector. The way out then consists
in promoting the global $N = (1,1)$ supersymmetry to a
superconformal $N = (1,1)$ supergravity.  We must therefore
associate to the two-dimensional graviton $h_{\alpha \beta}$ a
two-dimensional gravitino superpartner $\chi^\alpha$. Both fields
are non-dynamical: the graviton because of the Weyl and
diffeomorphism invariance, the gravitino because of super-Weyl and
local supersymmetry. Both will then contribute to the anomaly in
opposite directions, thus changing the critical dimension:
\begin{equation}
  D + \frac{D}{2} - 26 + 11 = 0 \Rightarrow D = 10 .
\end{equation}

\subsection{GSO projection and spectrum}

We can now consider the overall spectrum. Let us call
$\alpha_n^\mu$ and $\tilde \alpha_n^\mu$ the left- and
right-moving bosonic oscillators and $\beta_m^\mu $ and $\tilde
\beta_m^\mu$ the fermionic ones. Being
fermions, we have a choice on the boundary conditions that can be
either periodic or anti-periodic:
\begin{itemize}
  \item in the Neveu--Schwarz sector the conditions are anti-periodic and the
  $m$'s are half-integer;
  \item in the Ramond sector the conditions are periodic and the $m$'s are
  integer. In particular this allows for the presence of a zero-mode for which
  \begin{equation}
    \set{b_0^\mu , b_0^\nu } = \eta^{\mu \nu}.
\end{equation}
\end{itemize}
Depending on the sector, the ground state is respectively a
spacetime scalar or a Majorana--Weyl spinor.

Although the string spectrum is in general very constrained there
are ways to consistently remove or add (replace) full sectors
containing an infinite number of states. This is in general
possible in presence of discrete symmetries such as world-sheet
parity $\sigma \to - \sigma$, left fermionic number
$(-)^{F_\mathrm{L}}$ or right fermionic number
$(-)^{F_\mathrm{R}}$ (these are all $\setZ_2$ symmetries).  Any
projection must be shown to be consistent, \emph{i.e.} the
remaining states must form a closed set and if new states are
created they must be added in the form of new sectors. In
particular, the \textsc{gso} (Gliozzi--Scherk--Olive) projection
allows to consistently remove a large number of sectors, including
the tachyonic ones. Possible consistent projections are:
\begin{itemize}
\item type 0\textsc{a} and type 0\textsc{b} theories which have
  tachyons and no spacetime fermions
\item type \textsc{iia} and type \textsc{iib} theories without
  tachyons, with spacetime bosons and fermions, and 32 spacetime supercharges in
  ($N=2 $ supersymmetry in ten dimensions). In particular in \textsc{iia} the
  supersymmetry is non-chiral whereas it is chiral in \textsc{iib}.
\end{itemize}


\subsection{Open strings}
\label{cha:open-strings-1}

Let us go back for a moment to the simple situation of a bosonic
string described by the Nambu--Goto action:
\begin{equation}
  S = - T \int \di^2 \zeta \: \sqrt{\det (\hat g)} , \hspace{2em} \hat g_{\alpha \beta}
  = \partial_\alpha X^\mu \partial_\beta X^\nu \eta_{\mu \nu} .
\end{equation}
As it is usually the case, the variation $\delta S$ contains
boundary terms that we discarded in the previous derivation of the
Euler--Lagrange equations:
\begin{equation}
  \delta S_{\text{surface}} = -T \int_{\text{initial}}^{\text{final}} \left\{
  \left. \partial_\sigma X_\mu \delta X^\mu \right|_{\sigma = \pi} - \left.
  \partial_\sigma X_\mu \delta  X^\mu \right|_{\sigma = 0}
  \right\}.
\end{equation}
These must however vanish:
\begin{itemize}
\item for closed strings $X^\mu $ is periodic and $\left. X^\mu \right|_{\sigma
    = \pi} = \left. X^\mu \right|_{\sigma = 0}$ so the term is identically
    zero;
\item for open strings the two endpoints are independent and both the
  addends variations must vanish for an arbitrary variation.
\end{itemize}
A term of the form $\partial_\sigma X_\mu \delta X^\mu$ can vanish
in two ways, \emph{i.e.} each of the two factors can be zero. In the
general case $p + 1 $ directions are Neumann, \emph{i.e.} satisfy:
\begin{equation}
  \left. \partial_\sigma X_\mu \right|_{\sigma = \text{endpoint}} = 0 , \hspace{2em}
  \mu = 0, 1, \ldots, p ,
\end{equation}
and the $9 - p $ remaining are Dirichlet, \emph{i.e.} satisfy:
\begin{equation}
  \left. \delta X^\mu \right|_{\sigma = \text{endpoint}} = 0 , \hspace{2em} \mu = p +
  1, p + 2, \ldots, 9 .
\end{equation}
A special case
is given by $p = 9$, \emph{i.e.} when we have one D9-space-filling
brane. This is the so-called \emph{traditional open
  string}
\begin{equation}
  \left. \partial_\sigma X_\mu \right|_{\sigma = 0, \pi} = 0 , \hspace{2em} \mu = 0,
  1, \ldots, 9 ,
\end{equation}
and in terms of harmonic oscillators this condition implies
\begin{equation}
  \alpha^\mu_n = \tilde \alpha^\mu_n.
\end{equation}
This is as close as we can get to the closed string case; in
particular the vacuum state is unchanged: $\ket{ \mathbf{p}}$ for
the bosonic case and when adding fermions $\ket{\mathbf{p}}_{NS}$
is scalar and $\ket{ \mathbf{p}}_R$ is a Majorana-Weyl spinor.
Only one set of (bosonic or fermionic) oscillators act on these
vacua. The resulting two-dimensional theory is $N = 1 $ locally
supersymmetric, \emph{i.e.} has an $N=1$ surperconformal
supersymmetry (the difference with respect to the previous case is
that one of the supersymmetries is broken by the boundary
conditions). In some sense the spectrum is the ``square root'' of
the closed-string one.

The massless spectrum contains an NS bosonic part given by a spacetime
vector with $D-2 = 8$ physical components and a \textsc{r} fermionic
consisting in a spacetime Majorana-Weyl spinor.  They compose the
vector multiplet of type I $N=1$ supersymmetry algebra in ten
dimensions.

The present analysis
calls for some remarks:
\begin{itemize}
\item gravity is missing from the picture but this is not surprising
  since open strings by themselves do not constitute a consistent
  theory (they cannot merge to make closed strings). In other words a
  consistent theory has to be a collection of closed and open sectors;
\item there is an operation $\Omega : \sigma \to \pi - \sigma $ which changes the
  orientation of the string. States are even or odd with respect
  to this $\setZ_2$;
\item type IIA and IIB in ten dimensions do not have massless vectors
  in their perturbative spectrum. And when compactified give rise to
  Abelian $U(1)$'s;
\item if we take a bosonic left sector and a fermionic right one we
  obtain the heterotic theory with $N=1$ spacetime supersymmetry.
  Until the mid nineties heterotic string was the standard framework
  for phenomenology since it allows for non Abelian gauge groups: $E_8
  \times E_8$ or $SO(32)$.
\end{itemize}

The landscape of string theory has been drastically changed with
the appearance of D~branes. Nowadays most of the attempts to make
contact with phenomenology are based on type I and type II
theories in presence of D~branes.

Dirichlet~branes
are not rigid static objects. They have their own dynamics
(inherited by the attached open strings); moreover the end points
of open strings carry a $U(1)$ charge that couples to the $U(1)$
gauge field present on their spectrum. This $U(1)$ field
penetrates the D~brane and influences the dynamics. D~branes have
their own spectrum that demands a quantum mechanical description.
This can be obtained via the boundary conformal field theory
\textsc{bcft} of the open string. A semiclassical approximation is
nevertheless possible at low energies (the small parameter being
$\nicefrac{1}{\alpha^\prime}$ via the Dirac--Born--Infeld action).
Being extended objects, D~branes break the translational
invariance and therefore part of the spacetime supersymmetry,
which is an asset from the point of view of phenomenology.


\section{Strings in background fields}
\label{cha:strings-backgr-field}

\subsection{What is a background field?}
\label{sec:what-backgr-field}

In \textsc{qed} it is customary to study the classical field
created by a point-like charge and quantize the theory for test
particles in presence of such a background field. This is by its
very nature a semiclassical approximation because the background
satisfies the Maxwell equations. At the same time under certain
regimes the truly quantum nature of the system is bound to show
up. Contact between the classical relations and the quantum theory
is possible by describing the background fields as vacuum
expectation values of the field operators in coherent states.

A similar path can be followed in string theory where, although we
miss a full formalism similar to \textsc{qed}, as we have already
stressed above, we nevertheless have in flat space a
first-quantized version and a set of consistent and well-defined
perturbative rules.

Starting from the massless excitations we can build coherent
states and interpret them as background fields. The
(semi)classical interpretation is then possible under the
condition of not going too deep, \emph{i.e.} only in a
perturbative regime identified by the scales $\alpha^\prime$ and
$g_{\text{st}}$. We will only look at massless fields since the
massive ones would have a very short range and would take us away
from the classical background field approximation from the very
beginning.

To be more concrete we need to identify which elementary objects
are the sources for these backgrounds and what are the equations
that the semiclassical fields satisfy -- or equivalently which is
the low-energy effective action for the massless content of the
string spectrum. This means that the analysis of string theory in
non-trivial backgrounds will be important both for probing new
environments compatible with string dynamics potentially relevant
for phenomenology, and for going off-shell at least for the
massless excitations in some chosen regimes.

\subsection{Sources for antisymmetric fields}
\label{sec:sourc-antisymm-field}

To answer the first question,
\emph{i.e.} what are the elementary objects acting as sources for the
background fields, we must consider the massless component of the
spectrum. These are:
\begin{itemize}
\item the universal \textsc{ns} sector, $G_{\mu \nu}$, $B_{\mu \nu}$ and $\Phi$;
\item the open string gauge field $A_\mu$;
\item the \textsc{rr} forms which are $F\form{2} = \di A\form{1} $ and
  $F\form{4} = \di A\form{3}$ in type IIA and $F\form{1} = \di
  A\form{0}$, $F\form{3} = \di A\form{2}$ and $F\form{5} = \di
  A\form{4}$ in type IIB.
\end{itemize}
In the case of the electromagnetic field (a two-form) the action
is written as
\begin{equation}
  S = - \frac{1}{4 \kappa^2} \int \di^D x \: F_{\mu \nu} F^{\mu \nu} .
\end{equation}
The natural classical electric source for such a field is provided by
a point-like charge. The action is then given by:
\begin{multline}
  S^{\text{el}}_{\text{int}} = q \int_{\text{trajectory}} A = q \int A_\mu ( X(\tau)) \frac{\di X^\mu }{\di \tau } \di
  \tau = \\  = q \int \di \tau \left( \int \di^D x \delta^D ( X(\tau) - x ) A_\mu (x) \right)
  \frac{\di X^\mu }{\di \tau }
  = \int \di^D x \: A_\mu (x) j^\mu (x),
\end{multline}
where we have defined the electric current associated with the
point-like charge
\begin{equation}
  j^\mu (x) = \int \di \tau \: q \delta^D ( X(\tau) - x ) \frac{\di X^\mu }{\di \tau } .
\end{equation}

The dual magnetic field is a $\left( D - 2 \right)$-form $\ast F =
\tilde F$ generated by a $\left(D - 3 \right)$-form potential
$\tilde A$. This couples to a classical magnetic source, which is
an extended object with $\left(D -3 \right)$-dimensional
world-volume, or a $\left(D -4 \right)$~brane.
\begin{equation}
  S^{\text{mag}}_{\text{int}} = q_m \int_{\text{world-volume}} \tilde A =
  \int \di^D x \: \tilde A_{\mu_1 \mu_2 \ldots \mu_{D-3}} \tilde j^{\mu_1 \mu_2 \ldots
  \mu_{D-3}},
\end{equation}
where $\tilde j$ is the magnetic current associated with the $\left(D
  - 4 \right)$~brane.

For a more general $\left(p+2 \right)$-form $F\form{p+2}$ the
action reads:
\begin{equation}
  S = - \frac{1}{\left(p+2 \right)! \kappa^2} \int \di^D x \: F_{\mu_1 \mu_2 \ldots \mu_{p+2}} F^{\mu_1 \mu_2 \ldots \mu_{p+2}} .
\end{equation}
Generalizing the construction above, one sees that the natural
elementary classical electric sources for such a field are objects
with a $\left(p+1 \right)$-dimensional world-volume, \emph{i.e.}
$p$~branes:
\begin{equation}
  S_{\text{int}}^{\text{el}} = \int_{\text{world-volume}} A\form{p+1} = \int \di^D x \:
  A_{\mu_1 \mu_2 \ldots \mu_{p+1}} j^{\mu_1 \mu_2 \ldots
  \mu_{p+1}},
\end{equation}
whereas the magnetic sources are $\left(D-p-4 \right)$~branes:
\begin{equation}
  \int \tilde A\form{D-p-3} = \int \di^D x
  \tilde A_{\mu_1 \mu_2 \ldots \mu_{D-p-3}} \tilde j^{\mu_1 \mu_2 \ldots
  \mu_{D-p-3}}.
\end{equation}

In string theory, there exist a plethora of massless modes in
antisymmetric representations that can combine in coherent states
so to give the desired background. They are
\begin{itemize}
\item ordinary $U(1)$ gauge fields $A_\mu$ from the open string;
\item the antisymmetric \textsc{ns} tensor $H\form{3} = \di B\form{2}$
  which is electrically coupled to the fundamental string itself and
  magnetically coupled to a ($10 - 1 - 4 = 5$)~brane (the so-called
  NS5~brane);
\item in type IIA and IIB the various \textsc{rr} fields coupled as in
  Tab.~\ref{tab:fields-branes}.
\end{itemize}
\begin{table}
  \centering
  \begin{tabular}{|c|c|c|c|}
    \hline
     type & form & electric source & magnetic source  \\ \hline \hline
     IIA &  $F\form{2}$ & D0 & D6 \\ \hline
     IIA &  $F\form{4}$ & D2 & D4 \\ \hline \hline
     IIB &  $F\form{1}$ & D(-1) & D7 \\ \hline
     IIB &  $F\form{3}$ & D1 & D5 \\ \hline
     IIB &  $F\form{5}$ & D3 & D3 \\ \hline \hline
  \end{tabular}
  \caption{Field strengths and sources in type II strings}
  \label{tab:fields-branes}
\end{table}


\subsection{Generalizing the Polyakov action}
\label{sec:gener-poly-acti}

Up to this point we have only dealt in detail with strings
propagating in a trivial, flat background. This is not sufficient
because the string carries energy and electric charge with respect
to a 3-form field strength $H_{[3]}$. In other words it couples to
the metric $G_{\mu \nu }$, the dilaton $\Phi$ and the Kalb--Ramond
field $B_{\mu \nu }$ (\emph{i.e.} the two-form potential for
$H_{[3]}$).

A generalization of the Polyakov action
(Eq.~\eqref{eq:polyakov-flat}) is needed in order to capture these
more general situations. The most general action compatible with
two-dimensional renormalizability, Weyl and diffeomorphism
invariance reads:
\begin{equation}
  S = - \frac{1}{4\pi \alpha^\prime } \int \di^2 \zeta \: \sqrt{-h} \left( h^{\alpha \beta }
    \partial_\alpha X^\mu \partial_\beta X^\nu G_{\mu \nu } (X) +
    \epsilon^{\alpha \beta } \partial_\alpha X^\mu \partial_\beta X^\nu B_{\mu \nu } (X)
  - \alpha^\prime \Phi(X) R \right) .
\end{equation}
Although this action looks natural, it is not clear that it could
be obtained from first principles, based on a coherent-state
approach. A satisfying bottom-up approach should take into account
the fact that, when quantized around flat space, strings exhibit
quanta of $G_{\mu \nu }, B_{\mu \nu } $ and $\Phi$ fields
interacting in a very definite pattern. The way in which those
fields get further organized into coherent states is by no means
arbitrary and the way in which those, which we can read as
classical or semiclassical fields, couple to the string itself is
also definite.

This programme can be realized, with the following remarks though:
\begin{itemize}
\item as it is usually the case, there is an underlying assumption
  about the validity of the approximation which descends from the
  string topology expansion. This is crucial here since it becomes
  clear that the coupling $g_{\text{st}}$ is related to the
  \textsc{vev} of the dilaton as
  \begin{equation}
    g_{\text{st}} = \e^\lambda, \: \lambda = \braket{\Phi}
  \end{equation}
  and hence $\lambda $ is in general a function of the position: it is then
  possible to obtain a motion in the target space bridging a
  perturbative regime to a non-perturbative one;
\item the fields $G_{\mu \nu }, B_{\mu \nu} $ and $\Phi $ cannot be arbitrary
  since, at least in principle, they are coherent states of
  elementary strings states. In particular we shall demand that the same
  constraints of Weyl and diffeomorphism invariance are satisfied at
  the quantum level. In this case those requirements will not only
  impose a critical dimension but will translate into equations for
  the semiclassical fields;
\item $G_{\mu \nu }, B_{\mu \nu} $ and $\Phi $ appear as parameters in the
  two-dimensional theory, in general receive quantum corrections
  depending on the renormalization scheme and can be known exactly or
  only to some approximation in the parameter $\alpha^\prime $;
\item although $G_{\mu \nu }$ is a metric, its very geometric
  interpretation becomes questionable wherever the local curvature $R$
  is large with respect to $\nicefrac{1}{\alpha^\prime}$. This is
  due to the extension of the string ($\sim \sqrt{\alpha^\prime} $),
  which probes  $G_{\mu \nu }$.
\end{itemize}

\subsection{Equations of motion}
\label{sec:equations-motion}

The procedure for imposing Weyl invariance can be carried on in
different ways: compute amplitudes and demand the decoupling of
the conformal factor (Liouville mode) or compute the
$\beta$-functions and demand that they vanish. In any case it
remains technically involved. In general two outcomes are
possible:
\begin{itemize}
\item if from independent considerations we know that the model at hand
  is an exact two-dimensional \textsc{cft}, then we get
  a solution for all values of $\alpha^\prime$ (albeit a perturbative one);
\item only an $\alpha^\prime $-expansion is available.
\end{itemize}
In this latter case, for the bosonic \textsc{ns-ns} sector
we obtain:
\begin{subequations}
  \begin{align}
    \beta^G_{\mu \nu } &= \alpha^\prime R_{\mu \nu} - \frac{\alpha^\prime}{4} H_{\mu \lambda \rho}
    H_{\nu}^{\phantom{\nu}\lambda \rho} + 2 \alpha^\prime \nabla_\mu \nabla_\nu \Phi + \mathcal{O} (\alpha^{\prime 2} \partial^4) ,\\
    \beta^B_{\mu \nu } &= \alpha^\prime \nabla_\lambda H^\lambda_{\phantom{\lambda} \mu \nu } - 2 \alpha^\prime
    H^{\lambda}_{\phantom{\lambda}\mu \nu } \nabla_\lambda \Phi + \mathcal{O} (\alpha^{\prime 2} \partial^4) , \\
    \beta^\Phi &= \frac{D - D_{\text{cr}}}{6} - \frac{\alpha^\prime}{2} \nabla^2 \Phi + \alpha^\prime \nabla_\mu \Phi
    \nabla^\mu \Phi - \frac{\alpha^\prime}{24} H_{\mu \nu \rho} H^{\mu \nu \rho} + \mathcal{O} (\alpha^{\prime
      2} \partial^4) ,
  \end{align}
\end{subequations}
where $D_{\text{cr}}$ is the critical dimension that can be either
$D_{\text{cr}} = 10 $ or $D_{\text{cr}} = 26$.  It is a very
remarkable fact that those equations stem from the variation of an
effective action:
\begin{equation}
  S_{\text{eff}} = \frac{1}{2 \kappa_0^2} \int \di^{D_{\text{cr}}} x \: \sqrt{-G}
  \e^{-2 \Phi} \left[ R + 4 \partial_\mu \Phi \partial^\mu \Phi - \frac{1}{12} H_{\mu \nu \rho } H^{\mu \nu \rho }
    +  \mathcal{O} (\alpha^{\prime 2} \partial^4) \right].
\end{equation}
As we pointed out previously, this action provides both a way to
compute the allowed classical backgrounds where the strings
consistently propagate and an effective low-energy description of
the massless degrees of freedom. Obviously it receives
higher-order corrections. Fermions can also be included by
computing amplitudes that involve them. In this way the effective
action turns out to be a genuine supergravity action. This latter
aspect is extremely important because it provides a complete
description of how physics at Planck scale (described by string
theory) approximates at low-energies as usual field theory.

The above action is
written in the so-called \emph{string frame}, \emph{i.e.} with the
fields as they appear in the string sigma-model. This is not on the
other hand what one would get by a natural generalization of the
\textsc{gr} equations. For this reason an equivalent description is
usually given in the so-called \emph{Einstein frame} where the fields
are defined by
\begin{equation}
  \begin{cases}
    \Omega = \frac{2}{D-2} \left( \Phi_0 - \Phi  \right) \\
    \tilde G_{\mu \nu } = \e^{2\Omega} G_{\mu \nu} \\
    \tilde \Phi = \Phi - \Phi_0 \\
    \kappa = \kappa_0 \e^{\Phi_0}
  \end{cases}
\end{equation}
so that the action becomes
\begin{equation}
  S = \frac{1}{2 \kappa^2 } \int \di^D x \: \sqrt{- \tilde G}
  \left[ \tilde R - \frac{1}{6} \partial_\mu \tilde \Phi \partial^\mu \tilde \Phi - \frac{1}{12} \e^{-\nicefrac{\tilde \Phi}{6}} H_{\mu \nu \rho} H^{\mu \nu \rho} + \mathcal{O} (\alpha^{\prime 2} \partial^4 ) \right] .
\end{equation}

\subsection{Including Ramond--Ramond fields}
\label{sec:incl-ramond-ramond}

The fundamental string, which is the elementary object of the
perturbative approach leading to the equations of motion above,
does not couple to the Ramond--Ramond fields. In the case of the
three-form of type IIB this is because the string is not charged,
for all the other forms the very dimension of the string would not
allow for any couplings. This means in particular that there is no
straightforward way to incorporate the Ramond--Ramond backgrounds
in a sigma-model approach of perturbative string theory. This does
not mean that those fields are arbitrary. As a matter of fact they
still are coherent combinations of elementary string excitations
and they must satisfy some equations which can be computed in
terms of string amplitudes and then interpreted at low energies as
field-theory vertices. In this way, we obtain a low-energy
effective description for the \textsc{rr} fields that is given by
type I or type II supergravity in ten dimensions.

As an example we write the
bosonic sector for the type IIB string, in the Einstein frame:
\begin{equation}
\label{eq:IIB-action}
S_{\text{IIB}} = \frac{1}{2 \kappa^2} \int \di^{10} x \: \sqrt{-G} \left[ R -
  \frac{1}{2} \partial_\mu \Phi \partial^\mu \Phi - \frac{1}{12} \e^{-\Phi} H\form{3}^2 - \frac{1}{2} \e^{2\Phi} F\form{1}^2 - \frac{1}{12} \e^\Phi \tilde
  F\form{3}^2 - \frac{1}{48} F\form{5}^2 \right],
\end{equation}
where $\tilde F\form{3} = F\form{3} - A\form{0} \land H\form{3} $. Although
we will not expand on this, it is interesting to remark that the
couplings of the \textsc{ns} three-form and the \textsc{rr} three-form
are S-dual, \emph{i.e.} they get exchanged under $\Phi \to - \Phi$.

Similar expressions exist for type IIA and type I.


\section{Remarkable solutions}
\label{cha:remarkable-solutions}

\subsection{Brane-like solutions}
\label{sec:brane-like-solutions}

Solving the equations of motion is not in general an easy
task. A possible approach consists in making some ansatz and look for
special classes. In general, the backgrounds thus obtained will
receive higher-order $\alpha^\prime $ corrections. Some of them will turn out to
be exact (this is only possible in presence of pure \textsc{ns}
fields) and, at least in the near-horizon limit when the symmetries
are enhanced, the corresponding sigma model is an exact
\textsc{cft}.

The most simple ansatz is obtained for vanishing dilaton and
antisymmetric field strengths. In this case the equations reduce
to $R_{\mu \nu} = 0$, \emph{i.e.} the solution must be Ricci flat
as it is the case of Calabi--Yau manifolds. Less trivial
spacetimes are obtained by taking into account the other fields
that, from a general-relativity point of view, will contribute to
the energy momentum tensor of the system that will reflect in a
non-vanishing curvature. These solutions can then be interpreted
as backgrounds created by some classical, electrically or
magnetically charged D$p$~branes. In particular, if we allow for
the dilaton and a single $F\form{n}$ form the equations of motion
read
\begin{equation}
  \begin{cases}
    \displaystyle{R_{\mu \nu} = \frac{1}{2} \partial_\mu \Phi \partial^\mu \Phi + \frac{ \e^{\alpha \Phi}}{2
        \left(n-1 \right)!} \left( F_{\mu \mu_2 \ldots \mu_n }
        F_{\nu}^{\phantom{\nu}\mu_2 \ldots \mu_n} - \frac{n - 1}{8n} F_{\mu_1 \ldots \mu_n }
        F^{\mu_1 \ldots
          \mu_n} G_{\mu \nu} \right)} \\
    \displaystyle{ \nabla_\mu \left( \e^{\alpha \Phi} F^{\mu}_{\phantom{\mu}\mu_2 \ldots \mu_n} \right) = 0} \\
    \displaystyle{\triangle \Phi = \frac{\alpha }{2 n!} \e^{\alpha \Phi} F_{\mu_1 \ldots \mu_n } F^{\mu_1
        \ldots \mu_n}}.
  \end{cases}
\end{equation}
In the presence of a $p$~brane, the $SO(1,9)$ symmetry is broken
to $SO(1,p) \times SO(D-p-1)$, that is the $p+1 $ longitudinal
coordinates are separated from the $D - p -1$ transverse. The
natural ansatz is then
\begin{subequations}
  \begin{align}
    \di s^2 &= \e^{2 A(r)} \eta_{\alpha \beta} \di x^\alpha \di x^\beta + \e^{2 B(r)} \delta_{mn} \di y^m \di y^n \\
    \Phi &= \Phi (r),
  \end{align}
\end{subequations}
where $x^\alpha $ are the longitudinal coordinates, $y^m$ the transverse
and $r$ is the transverse radius
\begin{equation}
  r^2 = \sum_{m=p+1}^{D-1} y^m y^m.
\end{equation}
The ansatz for $F\form{n}$ depends on whether we choose an
electric or a magnetic coupling.

\paragraph{NS5~brane}

One of the most remarkable solutions is that obtained in type II
(A or B) in presence of an NS5~brane. If we choose a magnetic
coupling, corresponding to a solitonic-type brane, from the ansatz
above we obtain the solution
\begin{subequations}
  \begin{align}
    \di s^2 &= \frac{1}{H(r)^{\nicefrac{1}{4}}} \left( \di x^\alpha \di
      x^\beta \eta_{\alpha \beta } + H(r) \di y^m \di y^n \delta_{mn} \right) \\
    H\form{3} &= - \epsilon_{m_1 m_2 m_3}
    \partial_r H(r)\di y^{m_1} \land \di y^{m_2} \land \di y^{m_3} \\
    \e^\Phi &= \sqrt{H(r)},
  \end{align}
\end{subequations}
where the harmonic function $H(r)$ is
\begin{equation}
  H(r) = 1 + \frac{k}{r^2}
\end{equation}
with $k$ the NS5~brane charge. The initial ten-dimensional
Poincar\'e symmetry is clearly broken to Poincar\'e$_6 \times
SO(4) $.

If we consider the geometry close to the brane, \emph{i.e.} in the $r \to
0$ limit we get:
\begin{subequations}
  \begin{align}
    H(r) &\to \frac{k}{r^2} \\
    \e^\Phi &\to \frac{\sqrt{k}}{r} \\
    \di s^2 &\to \e^{-\nicefrac{\Phi }{2}} \left( \di x^\alpha \di x^\beta \eta_{\alpha \beta }
      + k \frac{\di r^2}{r^2} + k \di \Omega_3 \right),
  \end{align}
\end{subequations}
where $k \di \Omega_3$ is the line element for a three-sphere of
radius $\sqrt{k}$ \footnote{Notice that for $r\to 0$, the dilaton
diverges and we are naturally driven towards a strong-coupling
regime.}. Introducing the variable $z$ as
\begin{equation}
  r = \e^{-\nicefrac{z}{\sqrt{k}}},
\end{equation}
we observe that the background fields are those of the
\textsc{wzw} model on $SU(2)$ plus a linear dilaton. This is an
exact solution of string theory.

Other well-known applications of this kind of backgrounds can be
found in the so-called little string theory.

\paragraph{D5 in type IIB}
\label{sec:d5-type-iib}

For a D$5$~brane in type IIB we can choose a magnetic coupling
(getting once more a solitonic brane). In this case one can easily
verify that the solution is the same as before with the exchange
$\Phi \to - \Phi$ and $H\form{3}\to F\form{3}$. In other words the
two systems are S-dual.

No sigma-model interpretation is known, due to the presence of the
Ramond--Ramond field $F\form{3}$.

\paragraph{D3 in type IIB}
\label{sec:d3-type-iib}

Dirichlet~branes D3 in type IIB correspond to self-dual
$F\form{5}$ fluxes. In this case there is no dilaton and the
solution keeps a Poincar\'e$_4 \times SO(6)$ symmetry which can be
identified with an $\mathrm{AdS}_5 \times S^5$ geometry (the
symmetry is $SO(2,4) \times SO(6)$). Again this is not an exact
sigma-model but it has acquired an important role as the simplest
framework for implementing the holographic principle. According to
the latter, the theory in the bulk (the $\mathrm{AdS}_5$
background) is equivalent to the $N=4$ super-Yang--Mills theory
living on its border.

\paragraph{NS5-F1 system}
\label{sec:ns5-f1-system}

Another configuration possible in both type II theories is
obtained introducing a set of magnetically charged NS5~branes as
above and a set of electrically charged fundamental strings. In
general this solution will have a dilaton but in the $r\to0 $
limit the geometry is simply given by $\setR^4 \times
\mathrm{AdS}_3 \times S^3$ with an \textsc{ns} three-form on the
curved part and no dilaton. This can be easily identified with the
\textsc{wzw} model on the group $SL(2,\setR) \times SU (2)$ and it
is simple to show that supersymmetry or -- more strongly -- Weyl
invariance imposes the two curvatures to be equal in modulus.

Again it is possible to make an S-duality transformation and the
resulting system can be interpreted as a D5-D1 in type IIB for
which no sigma-model description is known.

\paragraph{M-theory solutions}
\label{sec:m-theory-solutions}

Ten-dimensional supergravity is not the most general theory
available. In fact, although less understood, a more general
theory can be formulated, admitting an eleven-dimensional $N=1$
\textsc{sugra} limit: the M-theory.

In this theory an $F\form{4}$ field is present, which couples
electrically to an M2~brane or magnetically to a M5~brane. The
simplest solution then consists in the M5~brane. This solution has
symmetry Poincar\'e$_6 \times SO(5) $ (it is in some sense a
generalization of the NS5 case above) and in the near-horizon
limit has geometry $\mathrm{AdS}_7 \times S^4$, with the four-form
flux proportional to the volume form of the sphere.

\subsection{Constant-curvature spaces}
\label{sec:const-curv-spac}

A final remark concerns the appearance of constant-curvature
spaces. The solutions above include only spheres or anti-de Sitter
spaces. This is not surprising as one can see by considering the
following heuristic argument. The Einstein equations in vacuum
with a cosmological constant read:
\begin{equation}
  R_{\mu \nu} - \frac{R}{2} G_{\mu \nu} + \Lambda G_{\mu \nu} = 0.
\end{equation}
Taking the trace one obtains
\begin{equation}
  R = \Lambda \frac{2D}{D-2}.
\end{equation}
The solution is therefore a constant-curvature space, which is
maximally symmetric: $\mathrm{AdS}_D, \mathrm{dS}_D, S^D$ or
$\mathrm{H}_D$, depending on the sign of $\Lambda$ and on the
signature.

The cosmological constant in the type of system we are considering
is effectively simulated by an $F^2$-term (see \emph{e.g.}
Eq.~\eqref{eq:IIB-action}) which is positive in the Euclidean case
and negative in the Minkowskian, leaving as only choices $S^n$ and
$\mathrm{AdS}_n$.

The search for de Sitter spaces in supergravity and string
theories has a long history. It was noticed long time ago that de
Sitter superalgebras lead very often to tachyonic spectra, which
are sources of instabilities. Despite that, vacua with positive
cosmological constant were found in gauged four-dimensional
supergravities. Whether those can be uplifted in some
higher-dimensional theory or string theory is however
questionable. This issue has recently attracted some attention,
but no clear construction for de Sitter-like solutions has emerged
in genuine string theory. Similar difficulties appear for
hyperbolic spaces, but any further discussion of this subject goes
beyond the scope of the present notes.

Let us finally stress that one should not exclude the possibility
that searching de Sitter vacua in string theory is of little
physical relevance. Firstly because de Sitter per se is not a
faithful description of the universe at any time. And, more
importantly, because the cosmological constant, as it is observed
today, is mainly an infrared effect that integrates all possible
scales and phase transitions. As such it is probably not captured
by a solution of a first-quantized string theory.


\nocite{Green:1987sp,Green:1987mn,Lust:1989tj,Polchinski:1996na,Townsend:1996xj,Kiritsis:1997hj,Dijkgraaf:1997ip,Sen:1998kr,Bachas:1998rg,Polchinski:1998rq,Polchinski:1998rr,D'Hoker:1988ta,Stelle:1998xg,Duff:1999rk,Aharony:1999ti,Forste:2001ah,Angelantonj:2002ct,Johnson:2003gi,Zwiebach:2004tj,Grana:2005jc}

\bigskip

\bibliography{Libri}

\ack

These notes are based on lectures delivered by Marios Petropoulos at
the \emph{ ``Corfu School and Workshops on High-Energy Physics''} held
in Corfu, Greece, in September 4 -- 26 2005, in the framework of the
European network QUEST. The authors acknowledge financial support by
the Agence Nationale pour la Recherche, France, contract 
05-BLAN-0079-01, and by the EU under
the contracts MEXT-CT-2003-509661, MRTN-CT-2004-005104 and
MRTN-CT-2004-503369. D.O. whishes to thank ENS in Paris and the
Arnold-Sommerfeld-Center for Theoretical Physics and the
Max-Planck-Institut für Physik in Munich where part of this work was
done and in particular D.Lüst, S.Reffert and S.Stieberger for useful
discussions. D.O. is supported by the European Commission FP6 RTN
programme MRTN-CT-2004-005104 and in part by the Belgian Federal
Science Policy Office through the Interuniversity Attraction Pole
P5/27 and in part by the ``FWO-Vlaanderen'' through project G.0428.06.

\end{document}